\documentclass[12pt]{article}

\usepackage{a4wide}
\usepackage[dvips]{graphicx}
\def\vp{{\mbox{\boldmath$p$}}}
\def\vz{{\mbox{\boldmath$0$}}}
\def\vQ{{\mbox{\boldmath$Q$}}}
\def\d{{-}}
\def\tlab{\mbox{$T_{\rm Lab}$}}
\def\tlabm{\mbox{$T_{\rm Lab}^{\rm max}$}}

\begin{document}

\begin{center}
{\bfseries
ON THE COVARIANT RELATIVISTIC SEPARABLE KERNEL}

\vskip 3mm

S.G. Bondarenko$^{\dag}$, V.V. Burov, E.P. Rogochaya,
and Y. Yanev

\vskip 5mm

{\small
{\it
JINR, 141980, Dubna, Moscow region, Russia
}
\\
$\dag$ {\it
E-mail: bondarenko@jinr.ru
}}
\end{center}

\vskip 5mm

\begin{center}
\begin{minipage}{150mm}
\centerline{\bf Abstract}
Two different methods of the covariant relativistic separable kernel
construction in the Bethe-Salpeter approach are considered.
One of them leads in the center-of-mass system of two particles
to the quasipotential
equation. The constructed 4-dimensional covariant functions
are used to reproduce neutron-proton phase shifts
for total angular momenta $J=0,1$. Obtained results are compared
with other model calculations.
\end{minipage}
\end{center}

\section{Introduction}
The relativistic
description  of the interaction between two
nucleons can be carried out using the Bethe-Salpeter (BS)
equation~\cite{BS} and different quasipotential approximations
to it~\cite{BSLT}. One of the ways to solve the BS equation
is using the separable presentation of its interaction kernel.
The BS formalism with such kernel have been successfully applied to
a covariant description of the elastic lepton-deuteron
scattering, deuteron electro- and photodisintegration, deep
inelastic scattering (DIS) of leptons on light nuclei.
It has also provided good
results for electromagnetic reactions with the deuteron.
The approach facilitates the analysis of the deuteron properties and
its comparison with nonrelativistic results
(see, review~\cite{obzor} and references therein).

The consideration of the deuteron
break up reaction demands the final state interaction between the
outgoing nucleons to be taken into account. If the separable
kernel of interaction is used it leads to the calculation of
integrals with separable functions. Usually they represent a
generalization of nonrelativistic separable functions~\cite{plessas}.
The presence of second- and higher-order poles in the integrand containing
these functions makes such calculation impossible.
In the paper~\cite{jetpl} we constructed the new
parametrization with functions suggested in~\cite{schwarz}, which do not
contain poles on the real axis in the relative energy complex plane, and apply
them to the description of the neutron-proton ($np$) scattering. Within
this approach we have achieved a good agreement with
experimental data in the region of high kinematic energies of the
nucleons $\tlab \sim 1.2$ GeV.
But as it was resumed in~\cite{jetpl} it is necessary to improve an agreement
with experimental data in a whole energy region up to
$\tlab<$ 3 GeV by increasing the rank of the separable kernel
of the nucleon-nucleon (NN) interaction.
This work is in progress.

Before that we investigate a case when the separable NN kernel
which represents a 3-dimensional function in the center-of-mass system (CMS).
The use of such functions leads to the approaches called quasipotential where the two-body Green's
function in the BS equation is replaced by a suitable function
which allows to reduce the 4-dimensional equation to 3-dimensional
one, for example, like in \cite{BSLT}.

In this work the one-rank separable kernels of NN interaction for
the partial states with total momenta $J=0,1$
are presented for functions in 4-dimensional and 3-dimensional forms.
Parameters of the kernels are fitted by the calculation of phase
shifts and low-energy observables (scattering length, effective
range, deuteron bound state energy) using the constructed $T$ matrix.

\section{Methods of a covariant relativistic generalization}

One of the common methods of a covariant relativistic generalization
of the Yamaguchi- and Tabakin-type functions is to replace 3-momentum
squared by 4-momentum squared:
\begin{eqnarray}
\vp^2 \to -p^2 = -p_0^2 + \vp^2.
\label{p2p}
\end{eqnarray}
This formal procedure converts 3-dimensional covariant functions
to 4-dimensional ones.

Another method is based on the introduction of the formal 4-vector $Q$ though
the relative $p$ and total $P$ 4-momenta of the two-body system
by the following relation:
\begin{eqnarray}
Q = p - \frac{P\cdot p}{s} P,
\label{Q2p}
\end{eqnarray}
with total momentum squared $s=P^2$.

Note that in CMS where $P=(\sqrt{s},\vz)$ the 4-vector $Q$
is defined by components $Q=(0,\vp)$ and thus
\begin{eqnarray}
\vp^2 = -Q^2
\label{p2Q}
\end{eqnarray}
can be formally converted to the Lorentz invariant.

Let us consider the methods described above in application to the
nonrelativistic Yamaguchi-type function
\begin{eqnarray}
g(|\vp|) = \frac{1}{\vp^2+\beta^2}.
\label{yams}
\end{eqnarray}

In the first case using the substitution (\ref{p2p}) we obtain the covariant function
in the form:
\begin{eqnarray}
g_p(p,P) = \frac{1}{-p^2+\beta^2} \stackrel{\rm CMS}{\longrightarrow}
\frac{1}{-p_0^2+\vp^2+\beta^2+i\varepsilon}.
\label{myams}
\end{eqnarray}

In the second case we use the relation (\ref{p2Q}) and obtain the function:
\begin{eqnarray}
g_Q(p,P) = \frac{1}{-Q^2+\beta^2} \stackrel{\rm CMS}{\longrightarrow}
\frac{1}{\vp^2+\beta^2}.
\label{myamQs}
\end{eqnarray}

The presented functions have rather different properties in the
relative energy $p_0$ complex plane in CMS. The function $g_p$ has two poles
on the real axis for $p_0$ at $\pm \sqrt{\vp^2+\beta^2} \mp i\varepsilon$ while the function $g_Q$
has no poles on it.

In practical calculations of the reactions with the high momentum transfer the $p_0$
integration can lead to singular expressions in functions $g_{p,Q}$
on $|\vp|$ or $\cos{\theta_{\vp}}$. Such problem can be easily solved by calculating
the $|\vp|$ or $\cos{\theta_{\vp}}$ principal value integral.
However, another form of functions $g_{p,Q}$ with odd powers
in a denominator leads to nonintegrable singularities. Therefore we introduce
functions $g_{p,Q}$ of type without poles on the real axis of the relative energy $p_0$
complex plane.
As an example of such function we introduce the covariant form factors
in the following form (see also the section 3 of the paper \cite{jetpl}):
\begin{eqnarray}
g_p(p,P) = \frac{p_c-p^2}{(p^2-\beta^2)^2+\alpha^4}
\stackrel{\rm CMS}{\longrightarrow}
\frac{p_c-p_0^2+\vp^2}{(p_0^2-\vp^2-\beta^2)^2+\alpha^4},
\label{myamsa}
\end{eqnarray}
and
in the second case we use the relation (\ref{p2Q}) and obtain the function
\begin{eqnarray}
g_Q(p,P) = \frac{p_c-Q^2}{(Q^2-\beta^2)^2+\alpha^4}
\stackrel{\rm CMS}{\longrightarrow}
\frac{p_c+\vp^2}{(\vp^2+\beta^2)^2+\alpha^4}.
\label{myamQsa}
\end{eqnarray}
We note that the function $g_Q$ still has no poles on the $p_0$ real axis while the
$g_p$ has poles at $p_0$:
$\pm \sqrt{\vp^2+\beta^2 + i\alpha^2},\quad
\pm \sqrt{\vp^2+\beta^2 - i\alpha^2}$.

The two methods of a covariant relativistic generalization described above
can be investigated by solving the Bethe-Salpeter equation
for specific partial states.

\section{Solution of the Bethe-Salpeter equation}

The details of the solution of the BS equation with the one-rank separable kernel
of interaction can be found in the paper \cite{jetpl}.
The value to be calculated is $h(s)$:
\begin{eqnarray}
h(s)=-\frac{i}{4\pi^3}\int dp_0\int|\vp|^2d|\vp|
S(p_0,|\vp|;s) {g(p_0,|\vp|)^2}
\label{hs}
\end{eqnarray}
with
\begin{eqnarray}
S(p_0,|\vp|;s) = \frac{1}{(\sqrt s/2-E_\vp+i\varepsilon)^2-p_0^2}
\label{sgr}
\end{eqnarray}
being the two-particle Green's function; here $E_\vp=\sqrt{\vp^2+m^2}$, $m$
is the nucleon mass.

To obtain the function $h(s)$ the 2-dimensional integral on $p_0$ and $|\vp|$
should be calculated. To perform the integration on $p_0$ the Cauchy theorem
is used. As it can be seen from Eq.(\ref{hs}) there are two types of singularities
in the $p_0$ complex plane: first ones are poles of the function $S(p_0,|\vp|;s)$:
\begin{eqnarray}
p^{(1,2)}_0 = \pm \sqrt{s}/2 \mp E_\vp \mp i\varepsilon,
\label{p0a1}
\end{eqnarray}
and other - poles of the function $g(p_0,|\vp|)$.

The function $g_p$ has four poles:
\begin{eqnarray}
p^{(3,4)}_0 = \pm \sqrt{\vp^2+\beta^2 + i\alpha^2},
\nonumber\\
p^{(5,6)}_0 = \pm \sqrt{\vp^2+\beta^2 - i\alpha^2},
\label{p0a2}
\end{eqnarray}
and to perform the $p_0$ integration residues in
three poles of Eqs.(\ref{p0a1}) and (\ref{p0a2}) should be calculated.
These calculations are performed analytically.

The function $g_Q$ has no poles on the $p_0$ real axis and therefore the only poles of
Eq.(\ref{p0a1}) should be taken into account. The result for $h(s)$
can be written as:
\begin{eqnarray}
h(s)=\frac{1}{2\pi^2}\int |\vp|^2d|\vp|
\frac{g_Q(0,|\vp|)^2}{\sqrt s-2E_\vp+i\varepsilon}.
\label{hsp}
\end{eqnarray}

This equation formally coincides with that could be obtained
within the Blankenbeckeler-Sugar-Logunov-Tavkhelidze (BSLT) approximation~\cite{BSLT}
which amounts to replacing the Green's function in Eq.(\ref{hs}) by
the expression
\begin{eqnarray}
S_{BSLT}(p_0,|\vp|;s) = -2\pi i (\sqrt s-2E_\vp+i\varepsilon)^{-1}  \delta(p_0).
\label{sgrbslt}
\end{eqnarray}

Although the solutions of the equation with functions $g_Q$ and in the BSLT
approximation coincide in CMS the difference becomes evident when the
reaction with the two-particle system is considered. In that case the
arguments of the function $g_Q$ are calculated with the help of
the Lorentz transformations in the system different from CMS.

\section{One-rank kernel}

We analyze two covariant relativistic generalizations
of the Yamaguchi form factors: modified Yamaguchi (MY) functions
and modified extended  Yamaguchi (MEY) functions.

\subsection{Modified Yamaguchi functions}

For the description of the chosen partial states we use the
following covariant expressions for both cases:
\begin{eqnarray}
g^{[S]}_p(p_0,\vp)=\frac{(p_{c1}-p_0^2+\vp^2)}{(p_0^2-\vp^2-\beta^2)^2+\alpha^4},
\label{simple_s}
\end{eqnarray}
\begin{eqnarray}
g^{[P]}_p(p_0,\vp)=\frac{\sqrt{-p_0^2+\vp^2}}{(p_0^2-\vp^2-\beta^2)^2+\alpha^4}.
\label{simple_p}
\end{eqnarray}
\begin{eqnarray}
g^{[S]}_Q(p_0,\vp)=g^{[S]}_p(Q_0,\vQ)
\stackrel{\rm CMS}{\longrightarrow}
\frac{(p_{c1}+\vp^2)}{(\vp^2+\beta^2)^2+\alpha^4},
\label{simple_sQ}
\end{eqnarray}
\begin{eqnarray}
g^{[P]}_Q(p_0,\vp)=g^{[P]}_p(Q_0,\vQ)
\stackrel{\rm CMS}{\longrightarrow}
\frac{|\vp|}{(\vp^2+\beta^2)^2+\alpha^4}.
\label{simple_pQ}
\end{eqnarray}
It should be noted that the 4-vector $Q$ is defined by the equation (\ref{Q2p}).
The functions are numerated by angular momenta $L=0([S]),1([P])$.
The numerator in $g^{[S]}_{p,Q}$ is introduced to compensate an
additional dimension in the denominator to provide the total
dimension as GeV$^{-2}$. This form was chosen because at
$p_{c1}=\beta^2$, $\alpha=0$ we get the function $g^{[S]}_{p,Q}$ in the
standard Yamaguchi form~\cite{manabe}. We prefer not to
consider the case with the square root in the denominator as it
is used in~\cite{schwarz} for $S$ waves to avoid the calculations
with cuts on the real axis in the $p_0$ complex plane.

\subsection{Modified extended Yamaguchi functions}

To extend the form of functions with increasing number of parameters
we introduce the following $g_{p,Q}$ functions:
\begin{equation}
g^{[S]}_p(p_0,\vp)=\frac{(p_{c1}-p_0^2+\vp^2)}{(p_0^2-\vp^2-\beta_1^2)^2+\alpha_1^4}
+\frac{C_{12}(p_0^2-\vp^2)(p_{c2}-p_0^2+\vp^2)^2}
{((p_0^2-\vp^2-\beta_2^2)^2+\alpha_2^4)^2},
\label{ext_s}
\end{equation}
\begin{equation}
g^{[P]}_p(p_0,\vp)=\frac{\sqrt{-p_0^2+\vp^2}}{(p_0^2-\vp^2-\beta_1^2)^2+\alpha_1^4}
+\frac{C_{12}\sqrt{(-p_0^2+\vp^2)^3}(p_{c3}-p_0^2+\vp^2)}
{((p_0^2-\vp^2-\beta_2^2)^2+\alpha_2^4)^2}.
\label{ext_p}
\end{equation}
\begin{equation}
g^{[S]}_Q(p_0,\vp)=g^{[S]}_p(Q_0,\vQ)
\stackrel{\rm CMS}{\longrightarrow}
\frac{(p_{c1}+\vp^2)}{(\vp^2+\beta_1^2)^2+\alpha_1^4}
-\frac{C_{12}\vp^2(p_{c2}+\vp^2)^2}
{((\vp^2+\beta_2^2)^2+\alpha_2^4)^2},
\label{ext_sQ}
\end{equation}
\begin{equation}
g^{[P]}_Q(p_0,\vp)=g^{[P]}_p(Q_0,\vQ)
\stackrel{\rm CMS}{\longrightarrow}
\frac{|\vp|}{(\vp^2+\beta_1^2)^2+\alpha_1^4}
+\frac{C_{12}|\vp|^3(p_{c3}+\vp^2)}
{((\vp^2+\beta_2^2)^2+\alpha_2^4)^2}.
\label{ext_pQ}
\end{equation}
The denominators with $p_{c1}, p_{c2}, p_{c3}$ are intended for a dimension
compensation as $p_{c1}$ for $g^{[S]}_{p,Q}$ in the previous case~(\ref{simple_s}), (\ref{simple_sQ}).

\begin{center}
\begin{table}
\caption{Parameters of the one{\d}rank kernels with MYQ functions.}
\centering
\begin{tabular}{lcc|lccc}
\hline\hline
                    & $^1S_0^+$ & $^3S_1^+$ &                     & $^1P_1^+$ & $^3P_0^+$ & $^3P_1^+$  \\
\tlabm              & 0.999\,GeV& 1.1\,GeV  &                     & 1.1\,GeV  & 0.999\,GeV& 0.999\,GeV \\
\hline
$\lambda$, GeV$^2$ & -351.73   & -0.048442 & $\lambda$, GeV$^4$ & 9.2481    &-172.92	  & 14.245     \\
$\beta$, GeV       &  1.8805   &  0.19973  & $\beta$, GeV       & 0.56446   & 0.72278   & 0.58444    \\
$\alpha$, GeV      &  0.24919  &  1.1891   & $\alpha$, GeV      & 0.2       & 0.68577   & 0.2        \\
$p_{c1}$, GeV$^2$  & -4.6069   & -332.95   &&&&\\
\hline
$s_0$, GeV$^2$     &  4.0279   &  4.2020   &                     &           & 3.8682    &            \\
\hline\hline
\end{tabular}\label{spar}
\end{table}
\end{center}

\begin{center}
\begin{table}
\caption{Parameters of the one{\d}rank kernels with MEYQ functions.}
\centering
\begin{tabular}{lcc|lccc}
\hline\hline
                       & $^1S_0^+$ & $^3S_1^+$&                        & $^1P_1^+$& $^3P_0^+$& $^3P_1^+$  \\
\tlabm                 & 0.999\,GeV& 1.1\,GeV &                        & 1.1\,GeV & 2.83\,GeV& 0.999\,GeV \\
\hline
$\lambda$, GeV$^2$   &-0.95380   &-16.852   & $\lambda$, GeV$^4$    & 0.02     &-197.47   &  0.031096  \\
$C_{12}$, GeV$^0$    & 3.2264    & 18.606   & $C_{12}$, GeV$^0$     & 0.15546  & 6.2551   & -0.071656  \\
$\beta_{1}$, GeV     & 0.10621   & 2.0098   & $\beta_{1}$, GeV      & 0.16353  & 0.67649  &  0.19350   \\
$\beta_{2}$, GeV     & 0.1       & 0.1      & $\beta_{2}$, GeV      & 0.44029  & 1.2192   &  0.65669   \\
$\alpha_{1}$, GeV    & 1.4281    & 0.20562  & $\alpha_{1}$, GeV     & 0.2      & 0.74382  &  0.2	  \\
$\alpha_{2}$, GeV    & 1.9491    & 1.1581   & $\alpha_{2}$, GeV     & 0.53818  & 1.1973   &  0.2	  \\
$p_{c1}$, GeV$^2$    & 27.424    & 132.22   & $p_{c3}$, GeV$^2$     & 103.92   &-2.8247   & -630.35    \\
$p_{c2}$, GeV$^2$    & 10.690    &-0.293    & &&&\\
\hline\hline
\end{tabular}\label{ext_param}
\end{table}
\end{center}

\begin{center}
\begin{table}
\caption{The low{\d}energy scattering parameters and binding energy for the
singlet $(s)$ and triplet $(t)$ $S$ waves.} \centering
\begin{tabular}{lcccccccc}
\hline\hline
           &$a_{s}$(Fm) & $r_{0s}$(Fm) &$a_{t}$(Fm) & $r_{0t}$(Fm) &$E_d$(MeV)     \\
\hline
MYQ        &-23.69      & 2.22         & 5.37       & 1.71         & 2.2246       \\
MEYQ       &-23.49      & 2.30         & 5.37       & 1.71         & 2.2246       \\
Experiment &-23.748(10) & 2.75(5)      & 5.424(4)   & 1.759(5)     & 2.224644(46) \\
\hline\hline
\end{tabular}\label{lep}
\end{table}
\end{center}

\begin{figure}[ht!]
\begin{center}
\includegraphics[width=130mm]{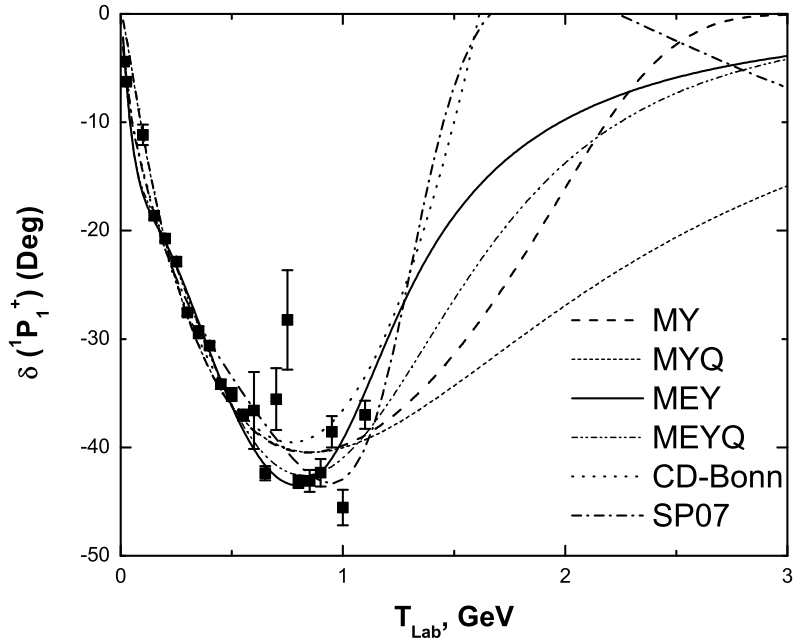}
\caption{{Phase shifts for the $^1P_1^+$ wave. Dash (short dash) line corresponds
to our parametrization with MY (MYQ) functions, eq.(\ref{simple_p})((\ref{simple_pQ})), respectively; the solid  (dash dot dot) line
illustrates the extended case MEY (MEYQ), eq.(\ref{ext_p})((\ref{ext_pQ})); the dotted and
dash-dotted lines describe the CD-Bonn and SP07 calculations,
correspondingly.}}
\label{1p1}
\end{center}
\end{figure}

\section{Discussion}

Using the $np$ scattering data we analyze the parameters of the separable kernels
as it was earlier presented in \cite{jetpl}.

The calculated parameters of the one-rank kernels
with MYQ and MEYQ functions and \tlabm\,
are listed in Tables~1 and 2.
In Table 3 the calculated
low-energy scattering parameters for $S$ waves are compared with
experimental data.

\begin{figure}[ht!]
\begin{center}
\includegraphics[width=130mm]{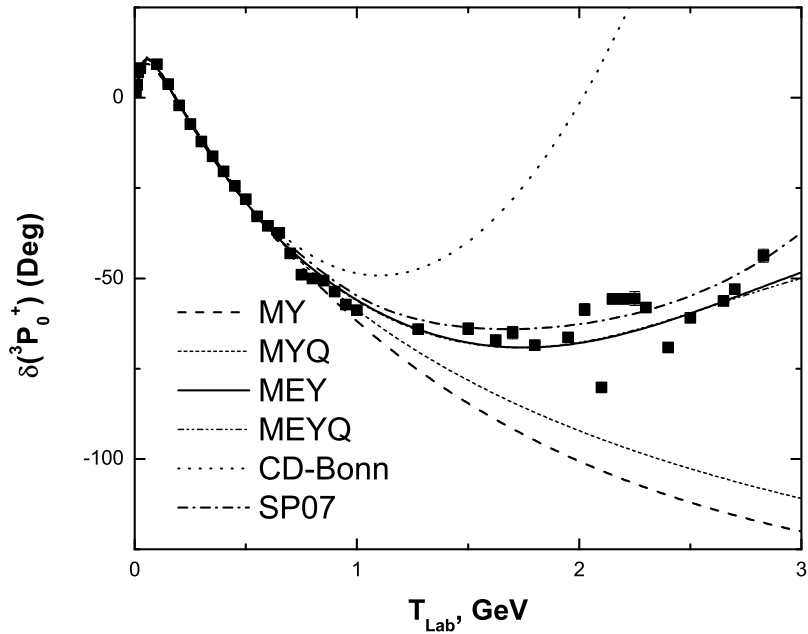}
\caption{{The same as in Fig.1 for the $^3P_0^+$ state.}}
\label{3p0}
\end{center}
\end{figure}

\begin{figure}[ht!]
\begin{center}
\includegraphics[width=130mm]{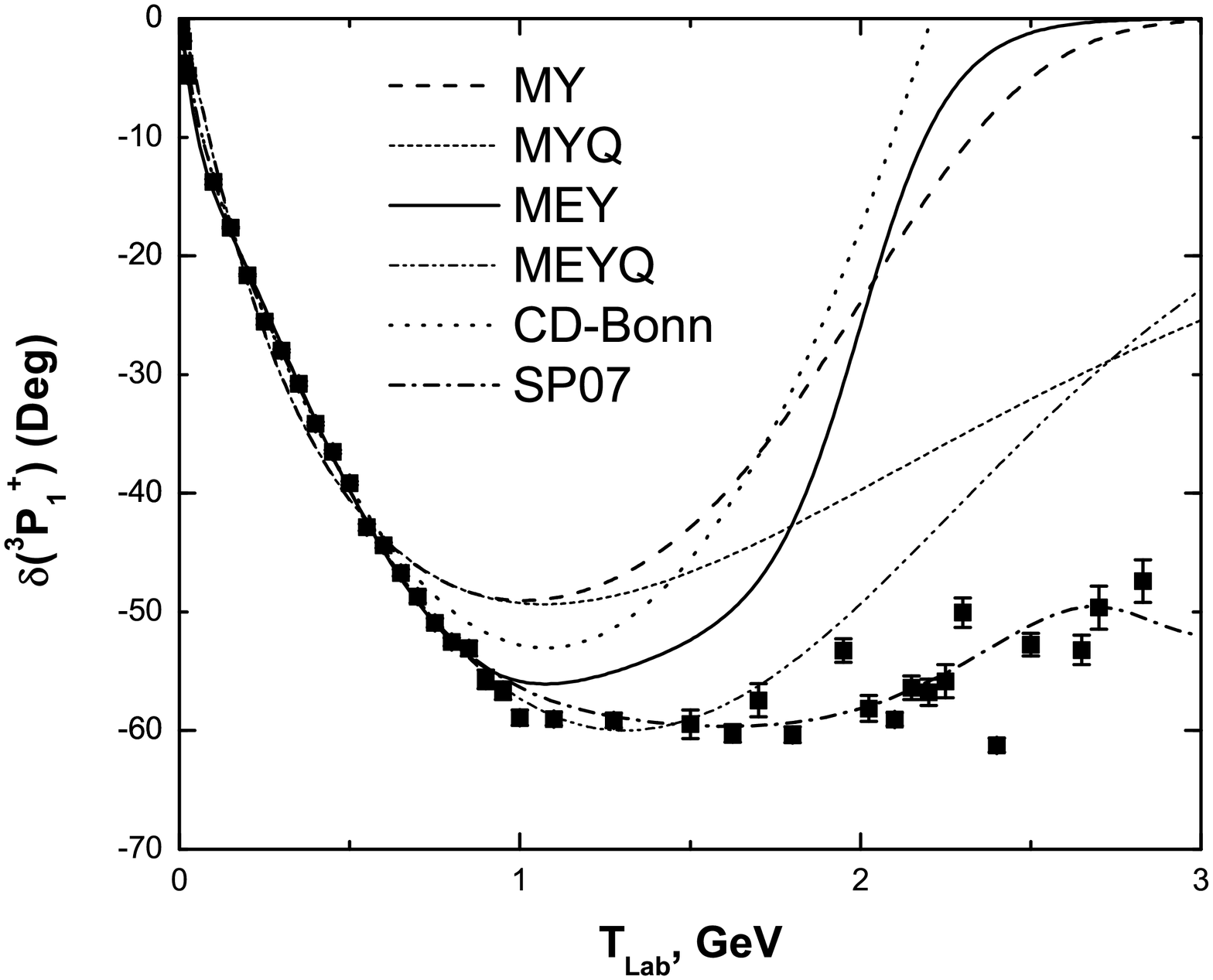}
\caption{{The same as in Fig.\ref{1p1} for the $^3P_1^+$ state.}}
\label{3p1}
\end{center}
\end{figure}

\begin{figure}[ht!]
\begin{center}
\includegraphics[width=130mm]{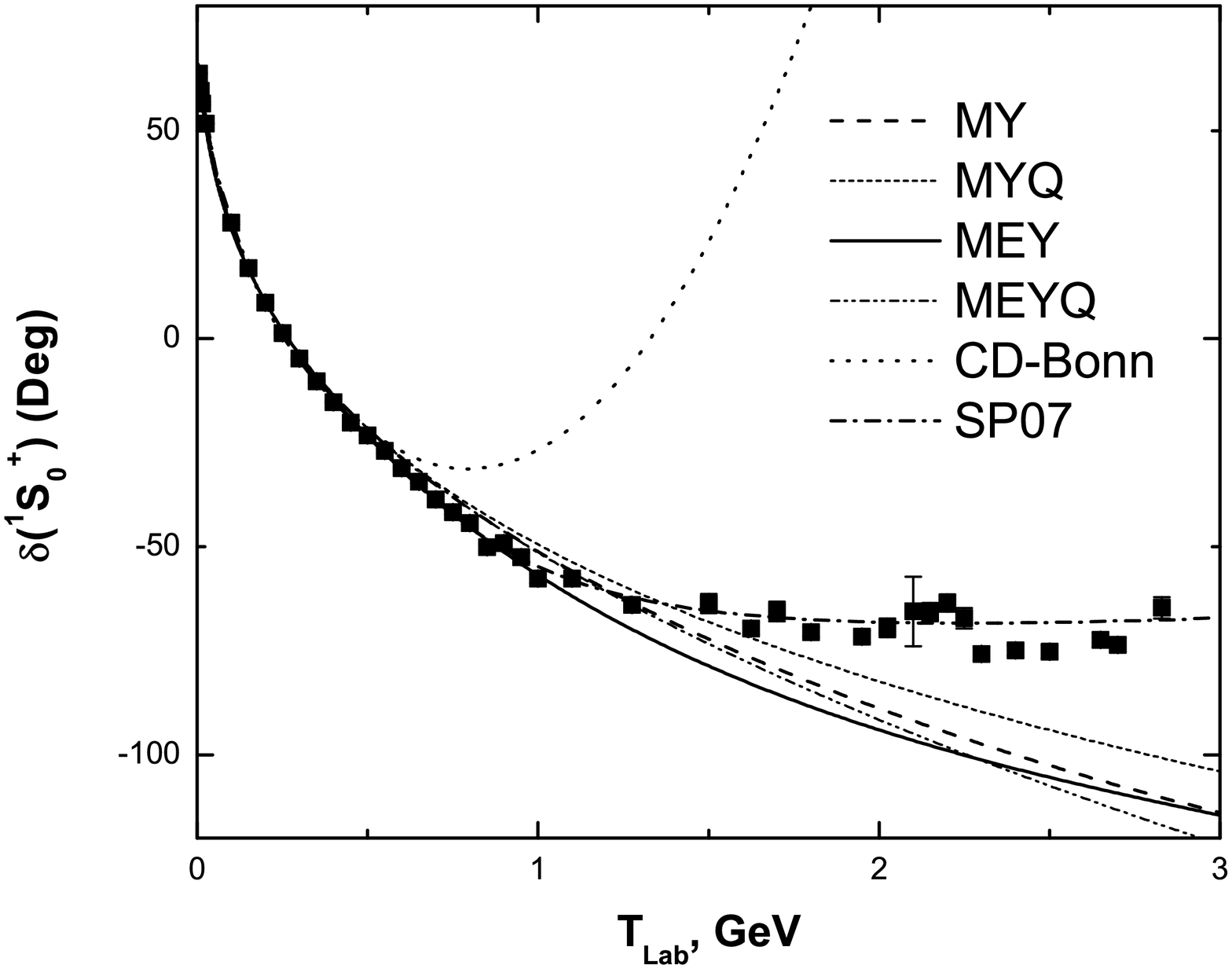}
\caption{{The same as in Fig.\ref{1p1} for the singlet partial state
$^1S_0^+$, but the description by MY (MYQ) functions is defined by eq.(\ref{simple_s})((\ref{simple_sQ}))
and extended one MEY(MEYQ) - by eq.(\ref{ext_s})((\ref{ext_sQ})).}} \label{1s0}
\end{center}
\end{figure}

\begin{figure}[ht!]
\begin{center}
\includegraphics[width=130mm]{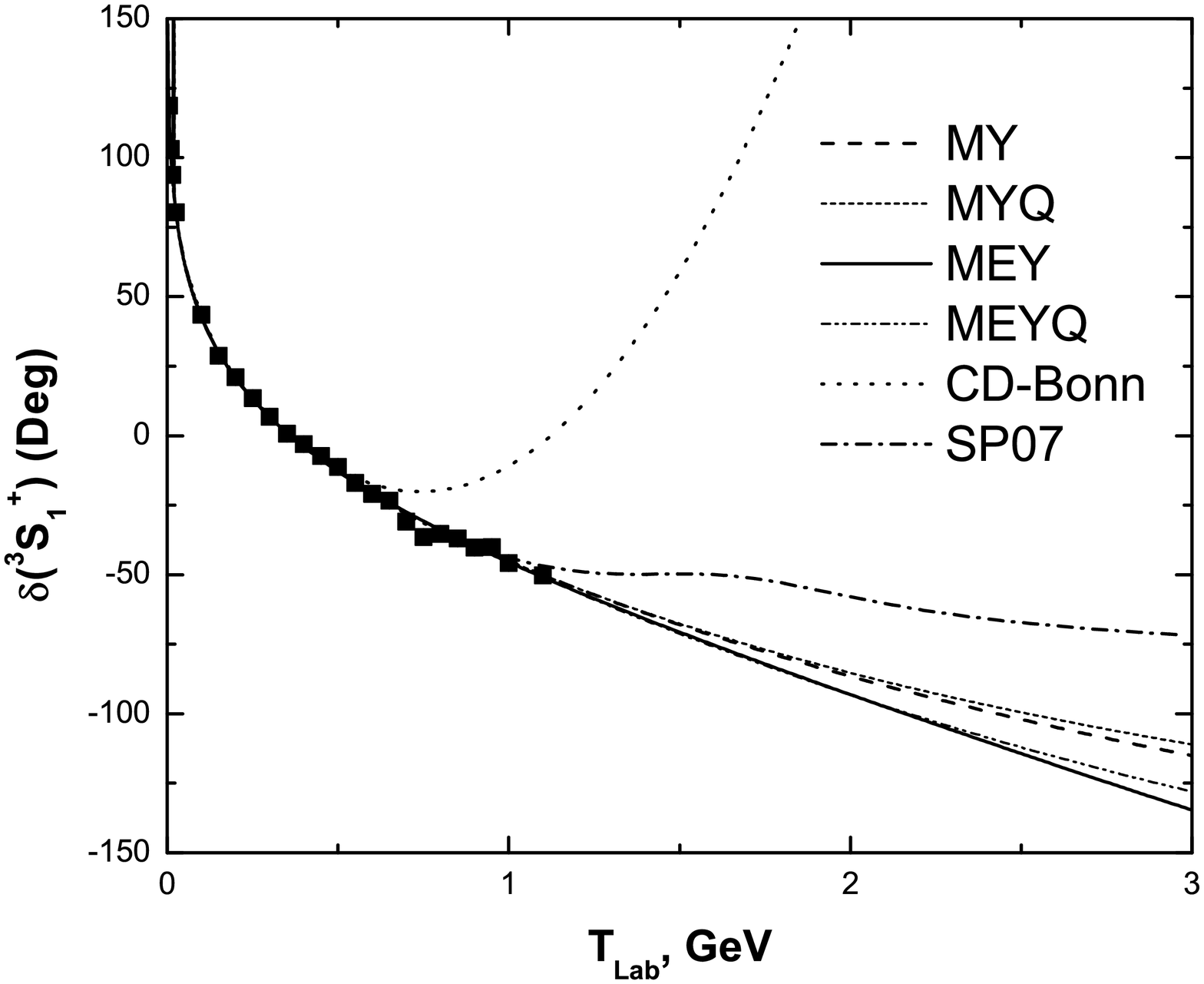}
\caption{{The same as in Fig.\ref{1s0} for the triplet partial state
$^3S_1^+$.}} \label{3s1}
\end{center}
\end{figure}

The calculated phase shifts are presented in Figs.\ref{1p1}-\ref{3s1} where
MY (MYQ) case is denoted by the dash (short dash) line, MEY (MEYQ) one - by the solid (dash dot dot) line.
Parameters for the first case (functions MY and MEY) have been presented in \cite{jetpl}.
The calculations with the nonrelativistic CD-Bonn \cite{bonn} (dotted line)
potential and the SP07 solution
\cite{SP07} (dash dot line) are included for comparison.

In Fig.\ref{1p1} we can see that all the calculations give
a reasonable agreement with experimental data up to the energy
\tlab $\sim 1.1$\,GeV but at larger energies their behavior becomes different
drastically. Note that the CD-Bonn and SP07 results for phase shifts even change
sign at $1.5< \tlab < 2.5$\,GeV. So we can stress that it is very desirable to obtain
more exact determination of experimental data for the $^1P_1^+$ channel in this energy range.
Now keeping in mind that the discrepancies between CD-Bonn, MEY, MEYQ and
SP07 calculations at $\tlab< 1.2$\,GeV are not so large we can trust our
calculations (in both cases) in this range of energies.

In Fig.\ref{3p0} CD-Bonn and MY (MYQ) calculations demonstrate an opposite behavior for $^3P_0^+$
phase shifts at $\tlab >1$\,GeV. The MEY (MEYQ results practically coincide with MEY) and SP07 calculations
give a very good agreement with experimental data in a wide range of energies
$\tlab < 3$\,GeV. Thus, both variants of the kernels of the NN interaction
in the $^3P_0^+$ channel are acceptable to be used in various relativistic calculations
of reactions.

Calculations for the $^3P_1^+$ channel
(see Fig.\ref{3p1}) show us that MY (MYQ), CD-Bonn, and MEY have a similar
behavior in a wide energy range, but are able to explain experimental data up
to $\tlab=$0.55, 0.6 and 1.2\,GeV, respectively.
It should be noted that MEYQ calculations describe phase shifts up to 1.5\,GeV.
The SP07 calculations give a suitable agreement with experimental data up
to 3\,GeV. Calculations for the cases 1 (MY, MEY) and 2 (MYQ, MEYQ) coincide up to 0.55\,GeV, but at larger energies
their behavior becomes significantly different.
So it is clear that to explain the data for the $^3P_1^+$ channel
in a wider energy range within our approach we need to increase the rank of the separable kernel
of the NN interaction.

Fig.\ref{1s0} demonstrates a close agreement with experimental data for CD-Bonn up to the energy
$\tlab < 0.6$\,GeV, MY and MEYQ - up to $\tlab < 1.3$\,GeV in the $^1S_0^+$
channel (note that these calculations give a very similar behavior of phase shifts
in a wide energy range up to 3\,GeV).
The MEY calculations give a very good description of phase shifts up to 1\,GeV. So we can conclude that in this case
as well as for $^3P_1^+$ we need to increase the rank of the kernel.
The SP07 calculations describe experimental data for the $^1S_0^+$ channel up to 3\,GeV.

Fig.\ref{3s1} for the $^3S_1^+$ wave shows us that CD-Bonn calculations agree
with experimental data up to 0.6\,GeV. 
All the other compared results demonstrate a
similar behavior in a wide energy range and explain known experimental data
up to $\tlab =$ 1.1\,GeV. It should be noted that calculations of
low-energy parameters with MY (MYQ) and MEY (MEYQ) form factors
give us a reasonable agreement with experimental data
(see \cite{jetpl} for the first case and Table 3 - for the second).

\section{Conclusion}
We present two new different parameterizations (4-dimensional and 3-dimensional in CMS)
of the separable kernel
of the NN interaction which are adopted for calculations at large energies.
As it was expected the MEY(MEYQ) functions give a better description
of the scattering data than the MY(MYQ) ones. Using the suggested MEY (MEYQ) form factors
the phase shifts are described in a whole range
of measured energies for the following partial states: $^1P_1^+$, $^3S_1^+$
($\tlab <$ 1.2\,GeV) and  $^3P_0^+$ ($\tlab <$ 3\,GeV).
The phase shifts for the $^1S_0^+$ and $^3P_1^+$ partial states can be described
in our approach up to $\tlab <$ 1.1\,GeV. To improve their agreement
with experimental data up to $\tlab <$ 3\,GeV it is necessary to increase
the rank of the separable kernel of the NN interaction. This work is in progress.
Finally we can say that 4-dimensional (Bethe-Salpeter type) as well as 3-dimensional
(quasipotential type) in CMS parameterizations give a good description of the considered
shift phases. Both of them are to be used in calculations of various electromagnetic
reactions.

\section*{Acknoledgements}

We wish to thank our collaborators A.\,A.~Goy,
K.\,Yu.~Kazakov and D.\,V.~Shulga
for fruitful discussions.

\newpage
\onecolumn

\end{document}